\PassOptionsToPackage{table,dvipsnames}{xcolor}

\documentclass{svproc}
\usepackage[percent]{overpic} 
\usepackage{hyperref}
\usepackage{tikz}
\usepackage{colortbl}      
\usepackage{tabularx}
\usepackage{graphicx}%
\usepackage{multirow}%
\usepackage{amsmath,amssymb,amsfonts}%
\usepackage{mathrsfs}%
\usepackage[title]{appendix}%
\usepackage{textcomp}%
\usepackage{manyfoot}%
\usepackage{booktabs}%
\usepackage{algorithm}%
\usepackage{algorithmicx}%
\usepackage{algpseudocode}%
\usepackage{listings}%
\usepackage{url}

\def\orcidID#1{\unskip$^{[#1]}$} 

\definecolor{comm0}{RGB}{255, 0, 0}   
\definecolor{comm1}{RGB}{65, 105, 225} 
\definecolor{comm2}{RGB}{170, 125, 201}  
\definecolor{comm3}{RGB}{255, 255, 0} 
\definecolor{comm4}{RGB}{0, 128, 128}   
\definecolor{comm5}{RGB}{216, 0, 115} 
\definecolor{comm6}{RGB}{0, 255, 0}   
\definecolor{comm7}{RGB}{14, 176, 41}   
\definecolor{comm8}{RGB}{164, 148, 247}   

\newcommand{\colorsquare}[1]{%
  \tikz[baseline=(sq.base)] \node[draw, fill=#1, minimum size=2ex] (sq) {};%
}

\begin{document}
\mainmatter              
\title{Kicking Politics: How Football Fan Communities Became Arenas for Political Influence}
\titlerunning{Kicking Politics}  
%
\author{Helen Paffard \and Diogo Pacheco\orcidID{0000-0002-8199-585X}
}
%
%

\institute{Computer Science Department, University of Exeter, UK\\
\email{\{hp533, d.pacheco\}@exeter.ac.uk}}

\maketitle              

\begin{abstract}
This paper investigates how political campaigns engaged UK football fan communities on Twitter in the aftermath of the Brexit Referendum (2016–2017). Football fandom, with its strong collective identities and tribal behaviours, offers fertile ground for political influence. Combining social network and content analysis, we examine how political discourse became embedded in football conversations. We show that a wide range of actors—including parties, media, activist groups, and pseudonymous influencers—mobilised support, provoked reactions, and shaped opinion within these communities. Through case studies of hashtag hijacking, embedded activism, and political “megaphones,” we illustrate how campaigns leveraged fan cultures to amplify political messages. Our findings highlight mechanisms of political influence in ostensibly non-political online spaces and point toward the development of a broader framework in future work.

\end{abstract}

\section{Introduction}\label{sec1}
The UK’s 2016 Brexit Referendum, in which 52\% voted to leave the European Union, was a watershed moment in British politics, reshaping the political landscape and deepening societal divisions. New polarised identities emerged that extended beyond traditional party affiliations, reflecting nationalist, globalist, and economic viewpoints~\cite{bastos_parametrizing_2018}. Similar patterns of polarisation were evident in the 2016 US presidential election, prompting scholars to draw parallels between the two events~\cite{ramswell_derision_2017}.

In this shifting landscape, online social media platforms such as Twitter (now X) have become central arenas for political influence, shaping opinion formation and election outcomes. While political candidates have used them to boost votes~\cite{kartsounidou_measuring_2023}, these platforms have also empowered non-traditional actors without formal political roles to advocate at scale~\cite{riedl_political_2023}, reaching audiences in ostensibly non-political spaces. One such space is football fandom.

Football is deeply embedded in British cultural identity~\cite{poulton_mediated_2004}, and online fan communities increasingly express identities that extend beyond sport. Their strong collective identities and tribal behaviours make them attractive targets for those seeking to shape public opinion. Political actors have recognised these communities as vehicles for influence, whether mobilising activism, promoting nationalist ideologies, or managing international reputations~\cite{moreau_life_2021}. The post-Brexit Referendum period (2016–2017), combined with the vibrancy of online football networks, provides a compelling context for studying such dynamics.

This paper addresses the question: \textbf{To what extent did political campaigns engage with and influence UK online football fan communities on Twitter during 2016–2017?} We examine exposure to political content, key discourse themes, influential actors, and the mechanisms through which political narratives were introduced and amplified. We show how political and football discourse became closely intertwined, with political parties, media outlets, activist groups, and pseudonymous influencers leveraging fan identities to mobilise support, provoke reactions, and shape opinion.


The study makes two main contributions:
\begin{enumerate}
\item A large-scale network analysis of approximately 95,000 tweets at the intersection of UK football and politics, revealing how political discourse was embedded within fan communities.
\item Three case studies of distinct influence mechanisms—hashtag hijacking, embedded activism, and political “megaphones”—demonstrating how campaigns exploited fan networks to circulate political messages.
\end{enumerate}

\section{Related Works}
\label{relatedwork}

Social media has reshaped political communication by enabling direct interaction between politicians and the public, while also empowering non-traditional influencers to shape debate~\cite{riedl_political_2023}. Some highlight benefits for mobilisation and democratic participation~\cite{lynn_calculated_2020}, while others emphasise risks of agenda-setting~\cite{mccombs_agenda-setting_1972}, selective amplification, and weakened editorial control~\cite{walsh_platform_2024}.

Football communication has similarly been transformed. Clubs use platforms to engage fans and stakeholders~\cite{guzman_towards_2021,romero-jara_more_2024}, while fans act as content creators, interacting directly with players and raising social or political issues~\cite{sanderson_social_2022}. Yet fan networks can also enable toxic behaviours such as hate speech~\cite{seijbel_expressing_2022}. As discourse increasingly extends beyond sport, online fan communities provide fertile ground for political influence.

SNA has been widely applied to electoral events, from inferring affiliations~\cite{mora-cantallops_influence_2021} and identifying influencers~\cite{lynn_calculated_2020} to tracking misinformation~\cite{bovet_influence_2019}. The \emph{network propaganda} model~\cite{benkler_network_2018} highlights how structures facilitate manipulation, while SNA has exposed tactics such as bots, sockpuppets, and astroturfing~\cite{pacheco_uncovering_2021,sela_signals_2025}. By contrast, applications to sport remain limited. Research has explored global fan sentiment during the 2014 World Cup using language similarity~\cite{pacheco2015football}, club rivalries in the UK and Brazil via attention dynamics~\cite{pacheco2016characterization}, and supporter diversity as a proxy for social disorganisation in the UK~\cite{pacheco2017using}. Guzmán et al.~\cite{guzman_towards_2021} further showed that Manchester United’s Twitter network contained sub-communities centred on politics and current affairs.

Content analysis complements SNA by offering thematic insights. In politics, it has been used to capture issue priorities, measure polarisation, and approximate public opinion~\cite{bastos_parametrizing_2018,cram_uk_nodate}. In football, applications include Brexit-related identity tensions~\cite{kearns_two_2023}, antisemitism~\cite{seijbel_expressing_2022}, and public health debates~\cite{moreau_life_2021}. Hybrid approaches that integrate SNA and content analysis combine structural and thematic insights, as seen in Brexit studies~\cite{lynn_calculated_2020} and Italian election networks~\cite{giglietto_it_2020}.

Despite these advances, research on political discourse within online UK football remains sparse. Most existing studies focus either on political events in general discourse or on football's commercial or cultural dimensions. Specific mechanisms of political influence within football-focused and other non-political spaces are underexplored. This study addresses these gaps by applying a hybrid SNA and content analysis within UK online football fan communities during a time of heightened political tension, providing a new perspective on spaces not traditionally seen as political arenas.

\section{Methods}

\begin{table}[t]
\centering
\caption{Descriptive summary of extracted politically-relevant tweets.}
\label{tab:dataset_summary}
\begin{tabularx}{\columnwidth}{@{} l X @{}}
\toprule
\textbf{Statistic} & \textbf{Value} \\
\midrule
Total tweets & 94,846 \\
Date range & 25 July 2016 - 26 October 2017 \\
Unique user accounts & 43,252 \\
Original tweets & 44\% (41,447) \\
Retweets & 35\% (33,018) \\
Quotes & 14\% (13,381) \\
Replies & 7\% (7,000) \\
Unique hashtags & 15,930 \\
Top political hashtags & brexit, ukip, trump, edl, nhs, bnp, maga, ge2017, ira, indyref2, putin, euref, snp, labour, donaldtrump \\
Political keywords & brexit, tory, corbyn, labour, ukip, libdem, snp, sturgeon \\
\bottomrule
\end{tabularx}
\end{table}

\subsection{Identifying Political Content}
We began with a dataset of 152.3M UK football-related tweets, collected via the Twitter Streaming API between July 2016 and October 2017 using accounts and hashtags associated with 44 Premier League and Championship clubs. From this corpus, 94,846 tweets containing political content were extracted for analysis.

Political tweets were identified using a curated list of 163 hashtags and 8 keywords derived from prior studies~\cite{cram_uk_nodate,lynn_calculated_2020,lee_using_2018}, iteratively refined to balance neutral and partisan terms. To minimise false positives, broad terms such as \emph{vote} were excluded. For context preservation, original tweets that political tweets had quoted or replied to were also included when present in the football dataset. By contrast, downstream replies that lacked curated terms were excluded, as they typically diverged from the initial topic and were fragmentary in the football-based sampling frame.

Table~\ref{tab:dataset_summary} summarises the dataset. Although political content accounted for fewer than 0.1\% of football tweets, prior work highlights how even low-volume political messaging can exert outsized influence through visibility, virality, and agenda-setting effects~\cite{huszr_algorithmic_2022}.

\subsection{Network Construction and Analysis}

We construct four types of networks and apply Louvain to detect communities, each serving different analytical aims. To analyse global network structures from the perspective of information diffusion potential and connectivity, standard global network properties of each were computed, including density, average degree, clustering coefficient, component structure, and average path length.

\textbf{Hashtag co-occurrence network (undirected)} --- Constructed to examine conversation topics reflected in hashtag usage. Nodes represent hashtags, and edges link hashtags appearing in the same tweet, weighted by co-occurrence frequency. Nodes are annotated as political, football, location, or other. 

\textbf{User interaction network (directed)} --- Constructed to examine engagement patterns and influencers. Nodes represent users, and edges capture retweet, reply, quote, or mention interactions, weighted by interaction frequency. To support the identification of influencers in the user interaction network, in-/out-degree, betweenness, and PageRank centralities are computed for each node. The top 20 users per metric and interaction type are annotated by actor type (politician, media outlet, football club, etc.), based on the user account profile description and account verification status. 

\textbf{User similarity network (undirected)} --- Constructed to identify affinities between users based on hashtag usage, even in the absence of direct interaction. Edges are derived from a user–hashtag matrix projected onto user–user space, with weights given by the cosine similarity of users’ hashtag usage vectors. Low-frequency hashtags ($<20$ uses) and single-tweet users are excluded. Only statistically significant edges ($p<0.05$, similarity $>0.45$) are retained. 

\textbf{Hashtag similarity network (undirected)} --- Constructed to identify topical associations between hashtags based on shared user engagement. The filtered user–hashtag matrix is projected onto hashtag–hashtag space, with edges weighted by the cosine similarity of user engagement vectors. Only statistically significant edges ($p<0.05$, similarity $>0.1$) are retained. 

These networks serve as the foundation for a higher-level analysis that links user and hashtag communities to broader discourse themes, described in the following subsection.

\subsection{From Communities to Discourse Themes}
To link user- and hashtag-level structures, we characterised user communities by their engagement with hashtag communities. Engagement strength was defined as the number of user–hashtag co-occurrences between members of a given user community and hashtags belonging to a given hashtag community, aggregated from the user–hashtag matrix. 

To capture broader thematic structure, hashtag communities were described by their proportional hashtag composition and subsequently clustered into \textit{overarching discourse themes} (football, political, UK location, other) using hierarchical clustering with Ward linkage. 

User community–hashtag community engagement patterns were then visualised with polar plots, which mapped the distribution of engagements and overlaid the higher-level discourse themes identified through clustering. This integrated analysis highlights how user groups connect with both hashtag clusters and the overarching themes of the discourse.


\begin{figure}[t]
    \centering
    \begin{minipage}[c]{0.48\textwidth}
        \vspace{0pt} 
        \includegraphics[trim=0.5cm 2cm 0.5cm 4cm, clip, width=\linewidth]{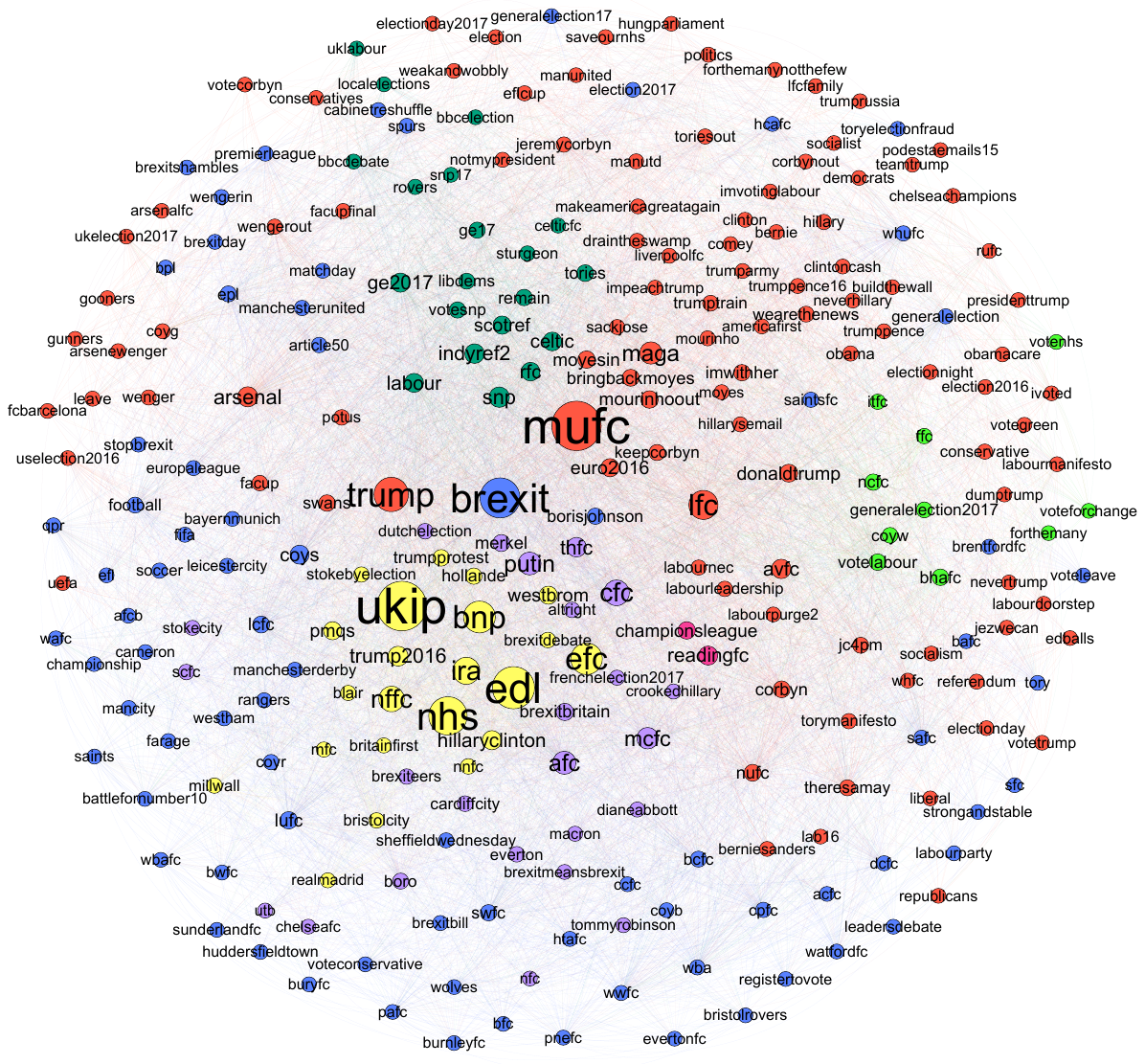}
    \end{minipage}%
    \hfill
    \begin{minipage}[c]{0.48\textwidth}
        \vspace{0pt} 
        \small
        \begin{tabular}{
            >{\raggedright\arraybackslash}p{5.6cm} 
            }
            \toprule
            \textbf{Discourse Topic Community}\\
            \midrule
            \colorsquare{comm0} Trump/MAGA; US \& UK Elections\\
            \colorsquare{comm1} Brexit-centric \\
            \colorsquare{comm2} Putin \& Mixed Themes \\
            \colorsquare{comm3} Far Right \& Nationalist \\
            \colorsquare{comm4} Scottish Independence; UK Election \\
            \colorsquare{comm5} Reading FC (Non-Political) \\
            \colorsquare{comm6} Labour Election Campaign \\
            \bottomrule
        \end{tabular}
    \end{minipage}
    \vspace{-12mm}
    \caption{\textbf{Hashtag Co-Occurrence Network and Discourse Topic Communities.} Network of football- and politics-related hashtags with $>25$ co-occurrences. Hashtags are interwoven, with some football hashtags strongly linked to specific discourse topics. Node colour indicates community; node size reflects weighted degree. Zoom recommended for detailed inspection.}

    \label{fig:hashtagnetwork}
\end{figure}

\begin{figure}[t]
    \centering
    \begin{overpic}[width=0.47\textwidth]{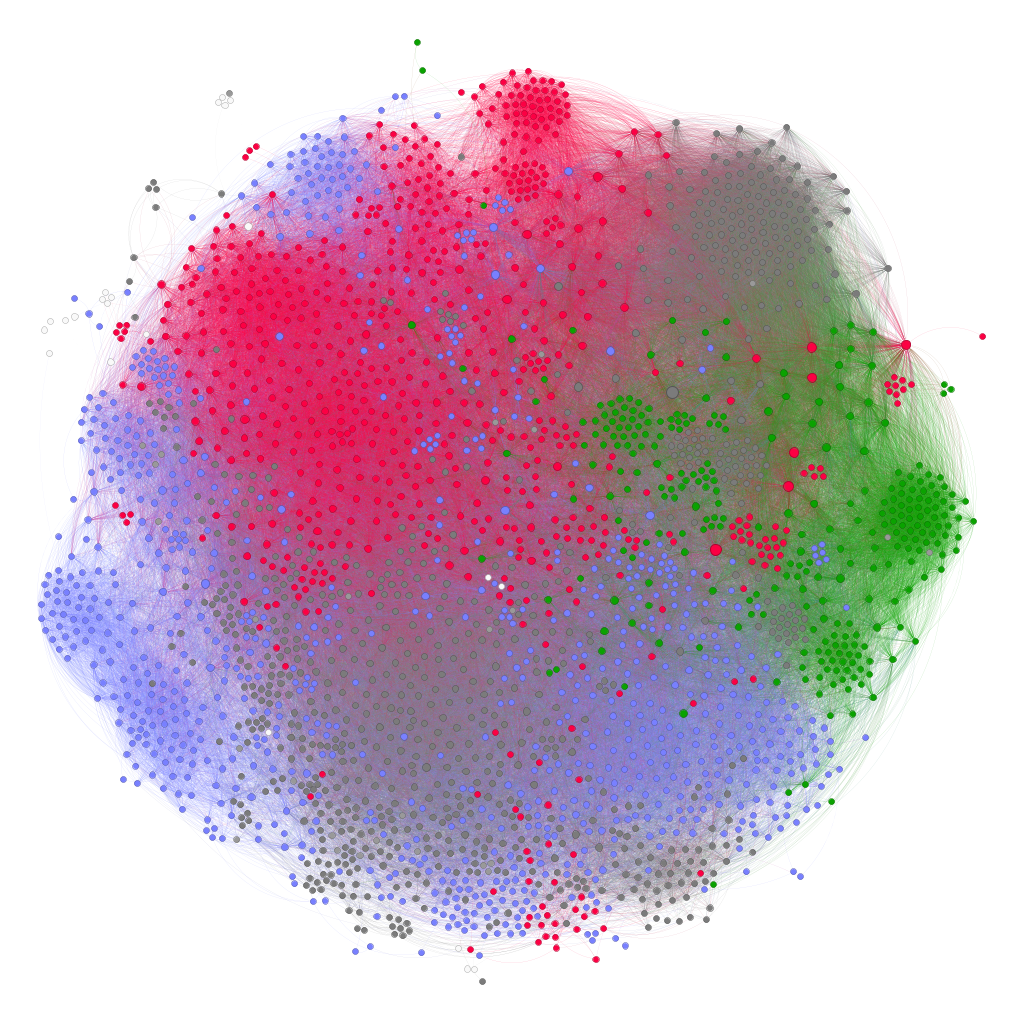}
        \put(4,90){\textbf{A}}  
    \end{overpic}
    \hfill
    \begin{overpic}[width=0.47\textwidth]{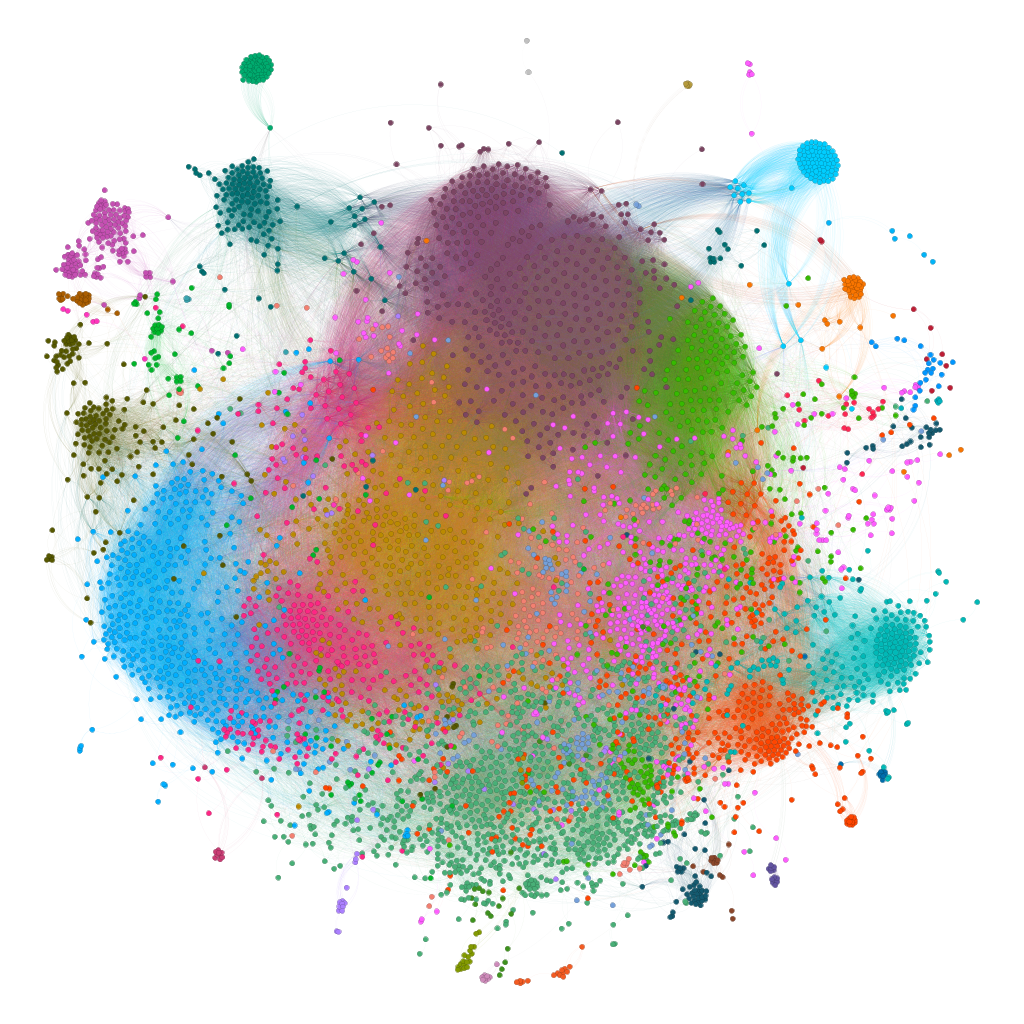}
        \put(4,90){\textbf{B}}
    \end{overpic}
    \vspace{-5mm}
    \caption{\textbf{Hashtag and User Similarity Networks.} One-mode projections of a user–hashtag bipartite matrix capturing topical alignment among users and hashtags. \textbf{(A)} Hashtag similarity network with nodes coloured by overarching discourse theme (\textbf{\textcolor{magenta}{Political}}, \textbf{\textcolor{comm8}{Football}}, \textbf{\textcolor{comm7}{UK}}, \textbf{\textcolor{gray}{General}}). \textbf{(B)} User similarity network with nodes coloured by user community.}
    \label{fig:similaritynetworksplus}
\end{figure}

\begin{figure}[ht]
    \centering

    \begin{minipage}[c]{0.43\textwidth}
        \vspace{0pt} 
        \includegraphics[trim=0 0 20 0, clip, width=\linewidth]{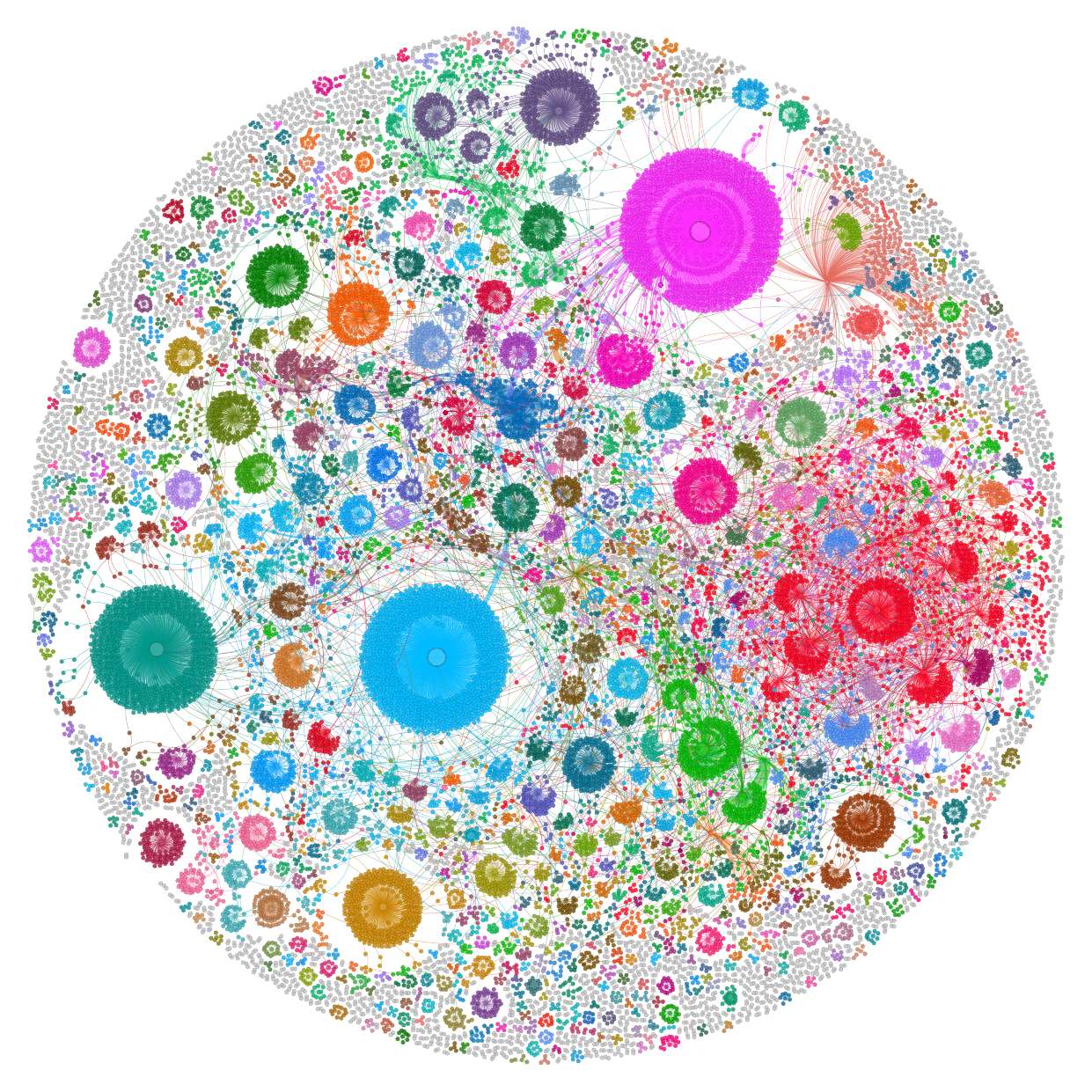}
    \end{minipage}%
    \hfill
    \begin{minipage}[c]{0.47\textwidth}
        \vspace{0pt} 
        \small
    \rowcolors{3}{gray!10}{white}
    \begin{tabular}{
        >{\raggedleft\arraybackslash}p{2.5cm} 
        >{\centering\arraybackslash}p{0.6cm} 
        >{\centering\arraybackslash}p{0.6cm} 
        >{\centering\arraybackslash}p{0.6cm} 
        >{\centering\arraybackslash}p{0.6cm} 
        }
        \toprule
        \textbf{Actor Type} & \multicolumn{4}{c}{\textbf{Interaction Type}} \\
        \cmidrule(lr){2-5}
        \rowcolor{white}
        & \textbf{RT} & \textbf{Q} & \textbf{R} & \textbf{M} \\
        \midrule
        Football Club        & -- & 1 & 11 & 15 \\
        Media                & 6  & 6 & 4  & 1 \\
        Politician/Party  & -- & 3 & 5  & 4  \\
        Fan News             & 3  & 5 & -- & -- \\
        Political User & 4  & 2 & -- & -- \\
        Activist Group       & 4  & 2 & -- & -- \\
        Football Fan         & 2  & -- & -- & -- \\
        Other                & 1  & 1 & -- & -- \\
        \bottomrule
    \end{tabular}
    \end{minipage}
    \vspace{-3mm}
    \caption{\textbf{User Retweet Network and Top Influencers.} 
    Node size represents PageRank; node colour indicates community. Zoom in for a detailed inspection.  \textit{Right:} Top 20 influencers by actor type and in-degree across interaction sub-networks (RT: retweet; Q: quote; R: reply; M: mention). Actor prominence varied by interaction type.}

    \label{fig:retweetnetwork}
\end{figure}

\section{Results}

\subsection{Topical Structure of Football-Political Discourse}

Content- and hashtag-based networks revealed dense connectivity, with giant components covering 98\% of nodes and high average degree, density, and clustering coefficients, indicating heavily overlapping discourse despite limited direct user interactions.

The hashtag co-occurrence network (15,930 hashtags; k-core $>25$) identified 264 tightly connected hashtags forming thematic clusters (Figure~\ref{fig:hashtagnetwork}). Community detection revealed political discourse—including partisan politics, electoral events, and cross-cutting issues such as Brexit—interwoven with football hashtags. Central political hashtags included \#Brexit, \#Trump, \#UKIP, and \#EDL\footnote{English Defence League, a far-right anti-Muslim extremist movement.}, while club hashtags such as \#MUFC, \#LFC, and \#Arsenal appeared across multiple political clusters. Scottish clubs (\#Celtic, \#RFC) clustered with Scottish independence topics, whereas other clubs (\#NFFC, \#EFC, \#CFC, \#THFC) aligned with far-right and nationalist discourse.

Hierarchical clustering of the hashtag similarity network condensed 27 communities into four overarching discourse themes: \textbf{Political}, \textbf{Football}, \textbf{UK}, and \textbf{General} (Figure~\ref{fig:similaritynetworksplus}A). Themes were distinct but highly interconnected, demonstrating that political and football conversations overlapped rather than existing in isolation.

\subsection{Mechanisms of Influence in User Networks}

The user similarity network identified 25 user communities (excluding those with $<10$ users), with 11 communities accounting for 90\% of users (Figure~\ref{fig:similaritynetworksplus}B). Mapping user communities to hashtag communities revealed several strongly political clusters, illustrating how content and user networks aligned to create influence pathways.

User interaction networks (retweets, quotes, replies, mentions) were fragmented, forming loosely connected clusters with low average degree, density, and clustering. The mention network, however, exhibited greater cohesion, with a giant component covering 87\% of nodes, longer average path length (4.05), and larger diameter (10), suggesting mentions connected more disparate users through elongated conversational clusters. Retweet networks displayed star-like cascades around influential accounts, indicating amplification of political content (Figure~\ref{fig:retweetnetwork}).

Actor prominence varied by interaction type. Activist groups and unverified individuals dominated retweets and quotes, leveraging these for message dissemination, while mentions and replies were led by verified institutions such as media outlets, football clubs, and politicians, which acted as conversational anchors. Betweenness centrality revealed divergent bridging roles: football institutions (@premierleague, @ManUtd, @Carabao\_Cup) connected cultural communities, while political actors and activist groups bridged clusters in retweets.

\subsection{Overlap Between Content and Users}

Together, the analyses indicate that politically- and football-focused communities co-existed and overlapped within a cohesive discourse space. Influence mechanisms ranged from one-to-many institutional broadcasting to multi-hub amplification by activist clusters, to conversational linking through mentions. These findings demonstrate how political discourse permeated football networks, with user communities and hashtags mutually reinforcing each other across multiple structural and thematic levels.

\subsection{Case Studies}
\vspace{-1mm}
\subsubsection{Case Study 1: Hashtag Hijacking} --- A pro-Trump influencer, ranked second by PageRank in the retweet network, hijacked \#MUFC alongside political hashtags \#tcot (Top Conservatives on Twitter) and the ironic \#LiberalsUnite to disseminate partisan content in a meme mocking Barack Obama: \textit{``Obama's greatest accomplishment is that he built the strongest Republican Party. Thank you, Obama. \#LiberalsUnite \#MUFC \#UNSC \#MAGA \#tcot''}. The tweet generated 1,413 retweets and 205 quotes.  

Manual retweeter profile analysis revealed that most retweeters were US Conservatives: 92\% of active retweeters ($>1$ tweet) belonged to US-politics-oriented user communities. Keyword-based profile analysis showed $<1\%$ of retweeter profiles had a football affiliation, compared with a mean of 22\% among typical \#MUFC users. Despite limited penetration into football audiences, a football news bot retweeted the post 11 times, illustrating how football club hashtag hijacks can help political messages infiltrate fan spaces, particularly when amplified by automated accounts.

\vspace{-5mm}
\subsubsection{Case Study 2: Hybrid Activism} 
\begin{figure}[t]
    \centering
    \includegraphics[width=0.48\textwidth]{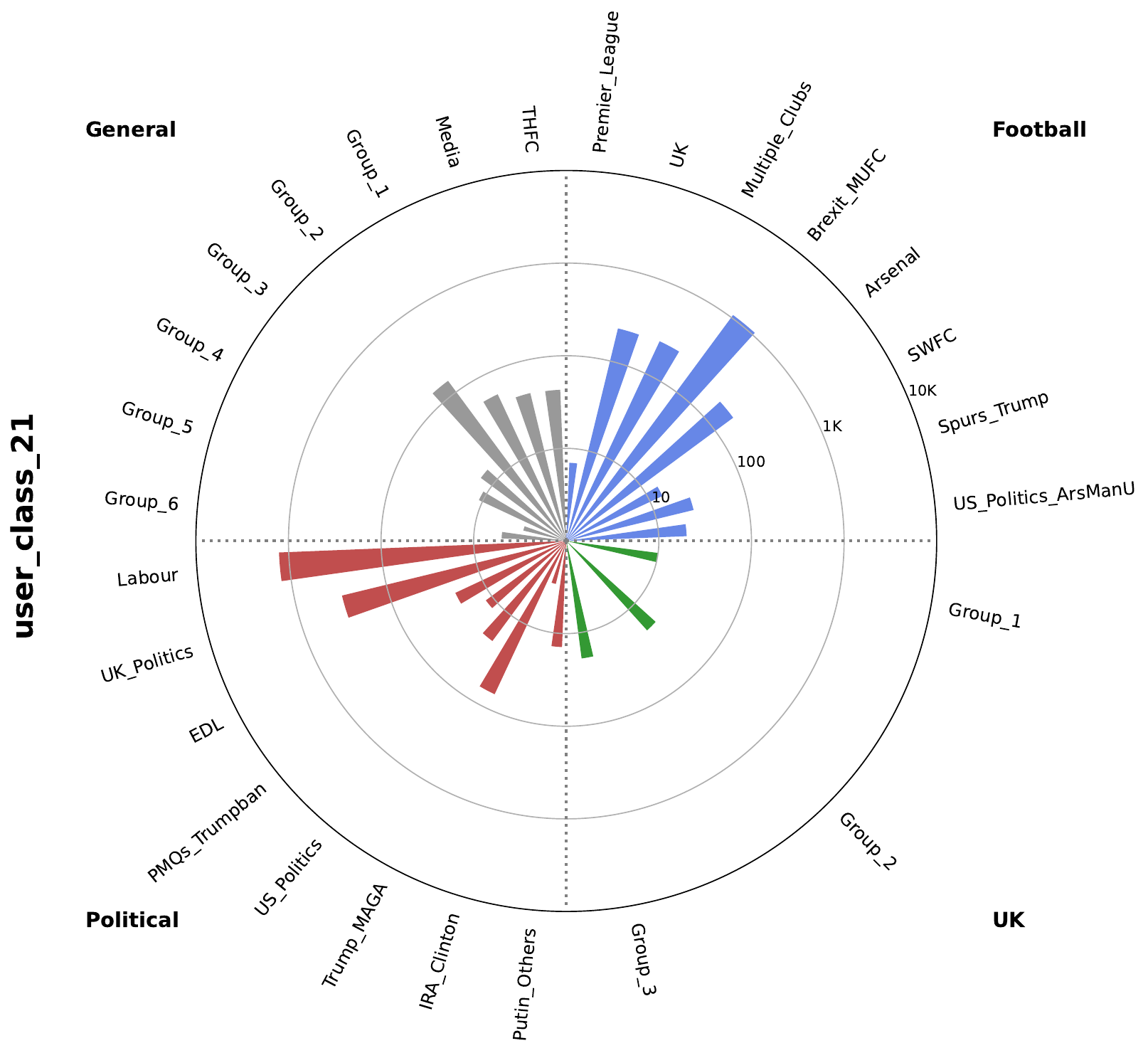}
    \hfill
    \includegraphics[width=0.48\textwidth]{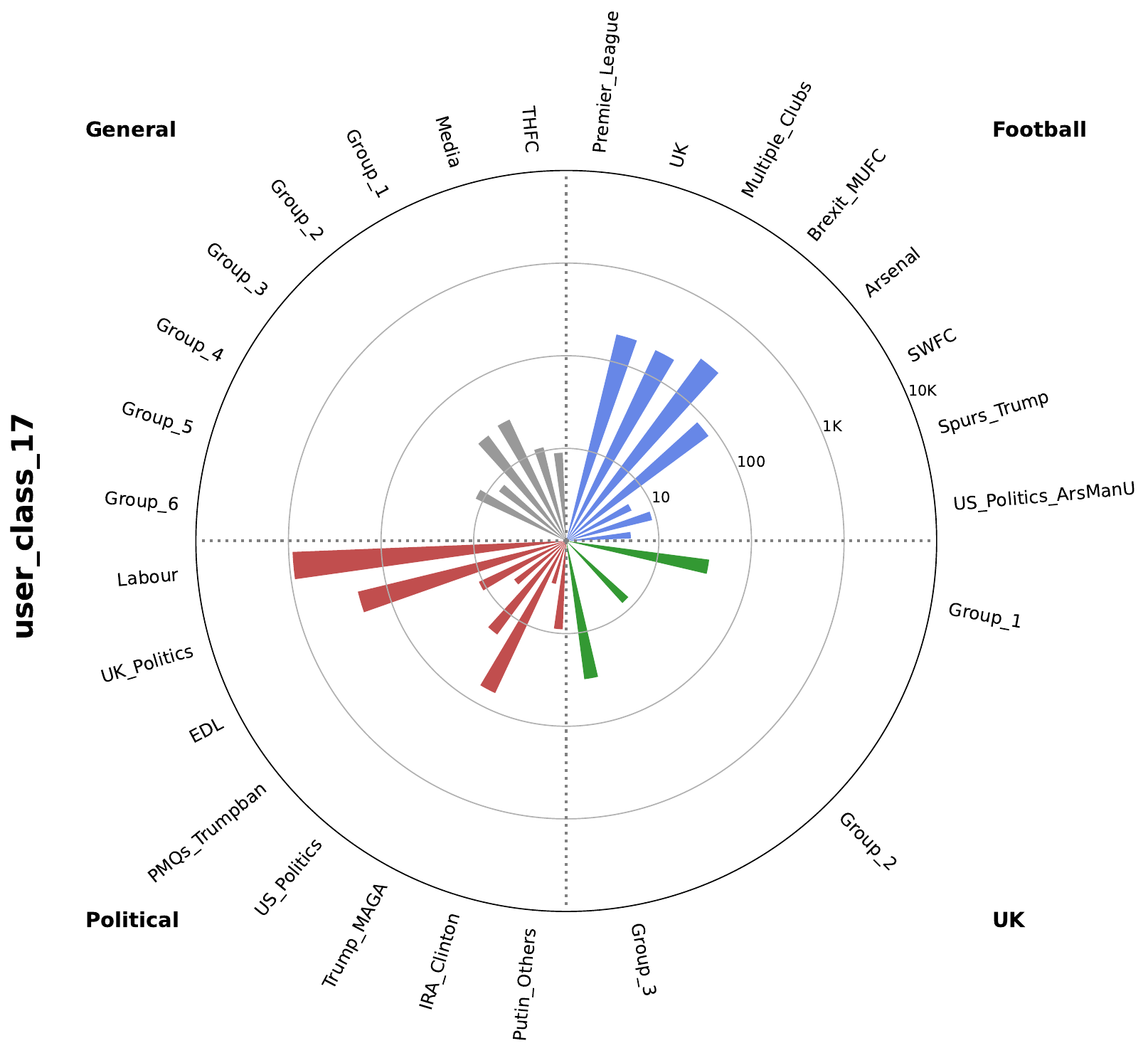}
    \vspace{-3mm}
    \caption{\textbf{Left-Leaning User Community Characterisation.} Polar plots display user community tweets per hashtag communities and overarching discourse theme. The two user communities centred on Labour Party politics and Jeremy Corbyn.}
    \label{fig:polarplots}
\end{figure}

--- During a football match on 7 May 2017, activists unveiled a banner supporting Labour leader Jeremy Corbyn alongside a campaign slogan\footnote{\href{https://bit.ly/hybrid_activism}{https://bit.ly/hybrid\_activism}}
. Online, a dense Labour-aligned cluster of 1,379 users within the retweet network, drawn from left-leaning user communities (characterised in Figure~\ref{fig:polarplots}), posted 619 tweets about the event. The largest cascades originated from two prominent activist accounts, self-described as ``\textit{ethical socialism}'' and ``\textit{helping to get Labour's General Election messages out and Jeremy Corbyn into No 10}.''  
This group exhibited significantly higher per-user engagement and retweet rates than the rest of the network (97\% vs. 65\%). The case illustrates the role of dense, ideologically aligned sub-networks in propagating political messages and exemplifies hybrid activism, where embedded political actors exploit fan network cohesion to amplify partisan messaging through authentic, bottom-up community engagement.

\vspace{-5mm}
\subsubsection{Case Study 3: Political Megaphone} --- Verified political actors, including Jeremy Corbyn and @UKLabour, ranked among the top 20 by in-degree across quote, reply, and mention sub-networks, despite minimal football-related posting: Corbyn tweeted once on Arsenal's FA Cup performance, and @UKLabour twice on grassroots football investment. Their influence derived from positional authority and frequent mentions (Corbyn 948, Labour 581), both supportive and oppositional, clustered along partisan lines. 

Supportive mentions concentrated within Labour-affiliated user communities, while oppositional mentions aggregated in broader UK politics communities. These accounts functioned as political megaphones, shaping discourse not through direct football engagement but by leveraging network visibility and distributed amplification. Supporters and opponents extended the reach of their messages, ensuring political narratives permeated fan communities even with minimal direct engagement.

\section{Discussion and Conclusion}

\paragraph{Key Findings in Relation to Research Aims.} 
This study examined how political campaigns engaged with UK online football fan communities on Twitter during 2016–2017. Although political content formed only a small fraction of the wider football discourse, individual tweets often achieved disproportionate reach. Interaction networks were fragmented yet connected by overlapping conversations, forming a cohesive discourse space even without direct user-to-user interaction. Diverse actors embedded political messages into everyday football talk, with evidence of hashtag hijacking, embedded activism, and amplification by bots, showing that football fandom had become a strategic terrain for political mobilisation.

Thematic analysis revealed political narratives spanning Brexit, the 2017 General Election, Jeremy Corbyn’s Labour Party, Scottish independence, Far Right messaging, and pro-Trump/MAGA content. These narratives were interwoven with football rather than isolated intrusions, with some cases crossing into mainstream media. Transatlantic linkages between Brexit and US populism were also visible, reflecting the broader political currents of the period.

While politicians and media commentators predictably set agendas, non-traditional actors such as activist groups, hyper-partisan influencers, and cyborg/bot accounts amplified messages within clustered communities. This aligns with network propaganda theory, which emphasises decentralised, mutually reinforcing influence dynamics over top-down broadcasting. Professional football players were largely absent from partisan exchanges, echoing evidence that their online activism typically addresses social issues rather than party politics.

Football clubs and organisations (e.g. Manchester United, Arsenal, Liverpool, Premier League, Carabao Cup) provided cultural anchors through their accounts and hashtags, linking otherwise disparate groups via their vast global fanbases. Club hashtags often appeared alongside political content, inadvertently acting as vectors for political messaging. Community structure suggested some fan tribalism, but less than expected at the outset.

Network analysis further showed how political narratives gained traction. For example, the dense Corbyn-supporting activist cluster spread messages effectively due to its integration into fan communities, whereas isolated hijack attempts had limited penetration. Influence was shaped less by message content or posting frequency than by structural position, underlining the importance of combining network and content analysis when examining influence strategies. 

\paragraph{Future Research Directions.} 
Our findings provide a foundation for a framework to detect and classify political influence in football and other cultural online spaces, informed by both network structures and content patterns. Future work should formalise this framework and assess its generalisability across broader domains. An interpretable approach integrating network and content analysis could support scalable detection for academic and policy purposes, while also underpinning supervised classifiers for systematic identification and risk assessment. Advancing this work will require bridging computational methods with political communication theory to capture evolving dynamics and cultural variation.

\paragraph{Limitations.} 
This study focuses on English-language Twitter content relating to Premier League and Championship clubs, offering a broad but incomplete view of UK online football fandom. It excludes platforms such as Facebook and Instagram, limiting demographic representativeness. Twitter’s user base skews towards politically engaged demographics~\cite{kartsounidou_measuring_2023}, so findings reflect politically active fans rather than the general public. Further limitations include: (1) reliance on hashtags and keywords, potentially overlooking indirect political expression; (2) incomplete coverage during key periods (June 2016 EU Referendum; partial June 2017 General Election); and (3) challenges in interpreting tone or intent, particularly sarcasm and humour, introducing unavoidable subjectivity even with manual review. 

\paragraph{Conclusion.} 
This study applied a multi-step political content extraction process and a mixed-methods approach, combining SNA and content analysis, to examine how political narratives penetrated non-political online football communities. It identified key influencers, transmission pathways, and amplification mechanisms, illustrated through case studies that highlighted specific influence strategies.

Findings confirm that political messaging was embedded within UK football fandom on Twitter during the observation period. Beyond documenting a historical case, the study lays the groundwork for a detection and classification framework that could be extended into AI-based systems for identifying influence strategies. Such work would require formalising network signatures but could inform governance frameworks and protective measures for cultural communities online. Safeguards for freedom of expression must remain central: excessive monitoring risks reinforcing dominant narratives and prompting self-censorship, echoing the ‘spiral of silence’.

The relevance of this research extends to 2025, as social media continues to shape political and sporting discourse, with X a key platform. The findings advance understanding of how political actors leverage pre-existing social structures, including fan networks and club affiliations, for mobilisation. State actors also increasingly use football for soft power projection, as seen in high-profile `sportswashing' examples such as the 2022 Qatar World Cup~\cite{hassan_qatar_nodate,ganji_rise_2023}. 
In an era of geopolitical instability, misinformation, and persistent polarisation~\cite{bovet_influence_2019}, understanding how divisive viewpoints spread is vital. 
By showing that football fan communities can act as conduits for political influence, this study contributes to wider efforts to build societal resilience against digital manipulation, protecting both fans and democratic discourse.

\subsection*{Declaration}

For the purpose of open access, the authors have applied a Creative Commons Attribution (CC BY) licence to any accepted manuscript version arising from this submission.

%
%
\bibliographystyle{abbrv}
\bibliography{references} 

@inproceedings{pacheco2017using,
  title={Using social media to assess neighborhood social disorganization: A case study in the United Kingdom},
  author={Pacheco, Diogo F and Oliveira, Marcos and Menezes, Ronaldo},
  booktitle={30th International Florida Artificial Intelligence Research Society Conference, FLAIRS 2017},
  pages={341--346},
  year={2017},
  organization={AAAI Press}
}

@inproceedings{pacheco2016characterization,
  title={Characterization of Football Supporters from Twitter Conversations},
  author={Pacheco, Diogo F and Pinheiro, Diego and De Lima-Neto, Fernando B and Ribeiro, Eraldo and Menezes, Ronaldo},
  booktitle={2016 IEEE/WIC/ACM International Conference on Web Intelligence (WI)},
  pages={169--176},
  year={2016},
  organization={IEEE Computer Society}
}

@inproceedings{pacheco2015football,
  title={Football conversations: what Twitter reveals about the 2014 World Cup},
  author={Pacheco, Diogo and de Lima Neto, Fernando Buarque and Moyano, Luis and Menezes, Ronaldo},
  booktitle={Brazilian Workshop on Social Network Analysis and Mining (BraSNAM)},
  year={2015},
  organization={SBC}
}

@article{bastos_parametrizing_2018,
	title = {Parametrizing {Brexit}: mapping {Twitter} political space to parliamentary constituencies},
	volume = {21},
	issn = {1369-118X, 1468-4462},
	shorttitle = {Parametrizing {Brexit}},
	url = {https://www.tandfonline.com/doi/full/10.1080/1369118X.2018.1433224},
	doi = {10.1080/1369118X.2018.1433224},
	language = {en},
	number = {7},
	urldate = {2025-03-01},
	journal = {Information, Communication \& Society},
	author = {Bastos, Marco and Mercea, Dan},
	month = jul,
	year = {2018},
	pages = {921--939},
}

@article{ramswell_derision_2017,
    title = {derision, division–decision: parallels between {Brexit} and the 2016 {US} presidential election},
    volume = {16},
    issn = {1680-4333, 1682-0983},
    url = {http://link.springer.com/10.1057/s41304-017-0112-0},
    doi = {10.1057/s41304-017-0112-0},
    language = {en},
    number = {2},
    urldate = {2025-04-29},
    journal = {European Political Science},
    author = {Ramswell, Prebble Q.},
    month = jun,
    year = {2017},
    pages = {217--232},
}

@article{kartsounidou_measuring_2023,
	title = {Measuring the impact of candidates’ tweets on their electoral results},
	volume = {20},
	issn = {1933-1681},
	url = {https://doi.org/10.1080/19331681.2022.2069181},
	doi = {10.1080/19331681.2022.2069181},
	number = {2},
	urldate = {2025-03-01},
	journal = {Journal of Information Technology \& Politics},
	author = {Kartsounidou, Evangelia and Papaxanthi, Dimitra and Andreadis, Ioannis},
	month = apr,
	year = {2023},
	pages = {154--168},
}

@article{riedl_political_2023,
	title = {Political {Influencers} on {Social} {Media}: {An} {Introduction}},
	volume = {9},
	issn = {2056-3051, 2056-3051},
	shorttitle = {Political {Influencers} on {Social} {Media}},
	url = {https://journals.sagepub.com/doi/10.1177/20563051231177938},
	doi = {10.1177/20563051231177938},
	language = {en},
	number = {2},
	urldate = {2025-03-08},
	journal = {Social Media + Society},
	author = {Riedl, Martin J. and Lukito, Josephine and Woolley, Samuel C.},
	month = apr,
	year = {2023},
	pages = {20563051231177938}}

@article{poulton_mediated_2004,
    title = {Mediated {Patriot} {Games}: {The} {Construction} and {Representation} of {National} {Identities} in the {British} {Television} {Production} of {Euro} ’96},
    volume = {39},
    issn = {1012-6902, 1461-7218},
    url = {https://journals.sagepub.com/doi/10.1177/1012690204049072},
    doi = {10.1177/1012690204049072},
    language = {en},
    number = {4},
    urldate = {2025-04-29},
    journal = {International Review for the Sociology of Sport},
    author = {Poulton, Emma},
    month = dec,
    year = {2004},
    pages = {437--455},
}

@article{moreau_life_2021,
	title = {“{Life} is more important than football”: {Comparative} analysis of {Tweets} and {Facebook} comments regarding the cancellation of the 2015 {African} {Cup} of {Nations} in {Morocco}},
	volume = {56},
	issn = {1012-6902},
	url = {https://doi.org/10.1177/1012690219899610},
	doi = {10.1177/1012690219899610},
	language = {en},
	number = {2},
	urldate = {2025-03-01},
	journal = {International Review for the Sociology of Sport},
	author = {Moreau, Nicolas and Roy, Melissa and Wilson, Andrew and Atlani Duault, Laetitia},
	month = mar,
	year = {2021},
	note = {Publisher: SAGE Publications Ltd},
	pages = {252--275},
}

@article{lynn_calculated_2020,
	title = {Calculated vs. {Ad} {Hoc} {Publics} in the \#{Brexit} {Discourse} on {Twitter} and the {Role} of {Business} {Actors}},
	volume = {11},
	url = {https://www.proquest.com/docview/2442453320/abstract/E1C34A6FC5C8432APQ/1},
	doi = {10.3390/info11090435},
	number = {9},
	urldate = {2025-03-01},
	journal = {Information},
	author = {Lynn, Theo and Rosati, Pierangelo and Nair, Binesh},
	year = {2020},
	pages = {435},
}

@article{mccombs_agenda-setting_1972,
    title = {The {Agenda}-{Setting} {Function} of {Mass} {Media}},
    volume = {36},
    issn = {0033-362X},
    url = {https://www.jstor.org/stable/2747787},
   number = {2},
    urldate = {2025-08-16},
    journal = {The Public Opinion Quarterly},
    author = {McCombs, Maxwell E. and Shaw, Donald L.},
    year = {1972},
    note = {Publisher: [Oxford University Press, American Association for Public Opinion Research]},
    pages = {176--187},
}

@article{guzman_towards_2021,
	title = {Towards understanding a football club’s social media network: an exploratory case study of {Manchester} {United}},
	volume = {49},
	copyright = {https://www.emerald.com/insight/site-policies},
	issn = {2398-6247, 2398-6247},
	url = {https://www.emerald.com/insight/content/doi/10.1108/IDD-08-2020-0106/full/html},
	doi = {10.1108/IDD-08-2020-0106},
	language = {en},
	number = {1},
	urldate = {2025-03-01},
	journal = {Information Discovery and Delivery},
	author = {Guzmán, Erick Méndez and Zhang, Ziqi and Ahmed, Wasim},
	month = feb,
	year = {2021},
	pages = {71--83},
}

@article{romero-jara_more_2024,
	title = {The more we post, the better? {A} comparative analysis of fan engagement on social media profiles of football leagues},
	volume = {25},
	copyright = {https://www.emerald.com/insight/site-policies},
	issn = {1464-6668},
	url = {https://www.emerald.com/insight/content/doi/10.1108/IJSMS-12-2023-0252/full/html},
	doi = {10.1108/IJSMS-12-2023-0252},
	language = {en},
	number = {3},
	urldate = {2025-03-01},
	journal = {International Journal of Sports Marketing and Sponsorship},
	author = {Romero-Jara, Edgar and Solanellas, Francesc and López-Carril, Samuel and Kolyperas, Dimitrios and Anagnostopoulos, Christos},
	month = jul,
	year = {2024},
	pages = {578--599},
}

@incollection{sanderson_social_2022,
	title = {Social {Media}, {Digital} {Technology}, and {National} {Identity} in {Sport}},
	copyright = {https://www.emerald.com/insight/site-policies},
	isbn = {978-1-80071-684-1},
	url = {https://www.emerald.com/insight/content/doi/10.1108/S1476-285420220000015013/full/html},
	language = {en},
	urldate = {2025-03-01},
	booktitle = {Research in the {Sociology} of {Sport}},
	publisher = {Emerald Publishing Limited},
	author = {Billings, Andrew C. and Anderson, Johnathan},
	editor = {Sanderson, Jimmy},
	month = apr,
	year = {2022},
	doi = {10.1108/S1476-285420220000015013},
	pages = {107--125}}

@article{seijbel_expressing_2022,
	title = {Expressing rivalry online: antisemitic rhetoric among {Dutch} football supporters on {Twitter}},
	volume = {23},
	issn = {1466-0970, 1743-9590},
	url = {https://www.tandfonline.com/doi/full/10.1080/14660970.2022.2109800},
	doi = {10.1080/14660970.2022.2109800},
	language = {en},
	number = {8},
	urldate = {2025-03-01},
	journal = {Soccer \& Society},
	author = {Seijbel, Jasmin and Van Sterkenburg, Jacco and Oonk, Gijsbert},
	month = nov,
	year = {2022},
	pages = {834--848}}

@article{mora-cantallops_influence_2021,
	title = {The influence of external political events on social networks: the case of the {Brexit} {Twitter} {Network}},
	volume = {12},
	issn = {1868-5137, 1868-5145},
	url = {https://link.springer.com/10.1007/s12652-019-01273-7},
	doi = {10.1007/s12652-019-01273-7},
	language = {en},
	number = {4},
	urldate = {2025-03-01},
	journal = {Journal of Ambient Intelligence and Humanized Computing},
	author = {Mora-Cantallops, Marçal and Sánchez-Alonso, Salvador and Visvizi, Anna},
	month = apr,
	year = {2021},
	pages = {4363--4375},
}

@article{bovet_influence_2019,
	title = {Influence of fake news in {Twitter} during the 2016 {US} presidential election},
	volume = {10},
	issn = {2041-1723},
	url = {https://www.nature.com/articles/s41467-018-07761-2},
	doi = {10.1038/s41467-018-07761-2},
	number = {1},
	urldate = {2025-03-01},
	journal = {Nature Communications},
	author = {Bovet, Alexandre and Makse, Hernán A.},
	month = jan,
	year = {2019},
	pages = {7}
	}

@book{benkler_network_2018,
    title = {Network {Propaganda}: {Manipulation}, {Disinformation}, and {Radicalization} in {American} {Politics}},
    isbn = {978-0-19-092362-4},
    url = {https://library.oapen.org/handle/20.500.12657/28351},
   language = {English},
    urldate = {2025-08-16},
    publisher = {Oxford University Press},
    author = {Benkler, Yochai and Farris, Robert and Roberts, Hal},
    year = {2018},
    doi = {10.1093/oso/9780190923624.001.0001},
    note = {Accepted: 2018-10-03 23:55},
}

@article{sela_signals_2025,
    title = {Signals of propaganda—{Detecting} and estimating political influences in information spread in social networks},
    volume = {20},
    url = {https://www.proquest.com/docview/3161750784/abstract/77541E3146847DFPQ/1},
    doi = {10.1371/journal.pone.0309688},
   number = {1},
    urldate = {2025-08-16},
    journal = {PLoS One},
    author = {Sela, Alon and Neter, Omer and Lohr, Václav and Cihelka, Petr and Wang, Fan and Zwilling, Moti and Sabou, John Phillip and Ulman, Miloš},
    month = jan,
    year = {2025},
    note = {Num Pages: e0309688
Place: San Francisco, United States
Publisher: Public Library of Science
Section: Research Article},
    pages = {e0309688},
}

@misc{pacheco_uncovering_2021,
    title = {Uncovering {Coordinated} {Networks} on {Social} {Media}: {Methods} and {Case} {Studies}},
    url = {http://arxiv.org/abs/2001.05658},
    doi = {10.48550/arXiv.2001.05658},
    urldate = {2025-08-16},
    publisher = {arXiv},
    author = {Pacheco, Diogo and Hui, Pik-Mai and Torres-Lugo, Christopher and Truong, Bao Tran and Flammini, Alessandro and Menczer, Filippo},
    month = apr,
    year = {2021},
    note = {arXiv:2001.05658 [cs]},
}

@article{hassan_qatar_nodate,
	title = {The {Qatar} {World} {Cup} and {Twitter} sentiment: {Unraveling} the interplay of soft power, public opinion, and media scrutiny,},
	volume = {59},
	doi = {https://doi.org/10.1177/10126902231218700},
	number = {5},
	journal = {International Review for the Sociology of Sport},
	author = {Hassan, A.A.M. and Wang, J.},
	pages = {679--704}
}

@article{kearns_two_2023,
	title = {Two {Brexits} on {Twitter}: {English} sporting identity and {Euro} 2016 as a metaphor for a divided {Britain}},
	volume = {24},
	issn = {1466-0970, 1743-9590},
	url = {https://www.tandfonline.com/doi/full/10.1080/14660970.2023.2192487},
	doi = {10.1080/14660970.2023.2192487},
	number = {8},
	urldate = {2025-03-01},
	journal = {Soccer \& Society},
	author = {Kearns, Colm and Sinclair, Gary and Lynn, Theo and Rosati, Pierangelo},
	month = nov,
	year = {2023},
	pages = {1091--1107},
}

@article{giglietto_it_2020,
	title = {It takes a village to manipulate the media: coordinated link sharing behavior during 2018 and 2019 {Italian} elections},
	volume = {23},
	issn = {1369-118X},
	url = {https://doi.org/10.1080/1369118X.2020.1739732},
	doi = {10.1080/1369118X.2020.1739732},
	number = {6},
	urldate = {2025-03-01},
	journal = {Information, Communication \& Society},
	author = {Giglietto, Fabio and Righetti, Nicola and Rossi, Luca and Marino, Giada},
	month = may,
	year = {2020},
	note = {Publisher: Routledge
\_eprint: https://doi.org/10.1080/1369118X.2020.1739732},
	pages = {867--891}}

@article{cram_uk_nodate,
    title = {{UK} {General} {Election} 2017: a {Twitter} {Analysis}},
    language = {en},
    author = {Cram, Laura and Llewellyn, Clare and Hill, Robin and Magdy, Walid},
}

@inproceedings{lee_using_2018,
    title = {Using {Twitter} {Hashtags} to {Gauge} {Real}-{Time} {Changes} in {Public} {Opinion}: {An} {Examination} of the 2016 {US} {Presidential} {Election}},
    isbn = {978-3-030-01159-8},
    urldate = {2025-08-16},
    booktitle = {Social {Informatics}},
    publisher = {Springer, Cham},
    author = {Lee, Hannah W.},
    year = {2018},
    note = {ISSN: 1611-3349},
    pages = {168--175},
}

@article{huszr_algorithmic_2022,
    title = {Algorithmic amplification of politics on {Twitter}},
    volume = {119},
    issn = {0027-8424},
    url = {https://www.jstor.org/stable/27112824},
    number = {1},
    urldate = {2025-08-16},
    journal = {Proceedings of the National Academy of Sciences of the United States of America},
    author = {Huszar, Ferenc and Ktena, Sofia Ira and O?Brien, Conor and Belli, Luca and Schlaikjer, Andrew and Hardt, Moritz},
    year = {2022},
    note = {Publisher: National Academy of Sciences},
    pages = {1--6},
}

@article{ganji_rise_2023,
	title = {The {Rise} of {Sportswashing}},
	volume = {34},
	issn = {1086-3214},
	url = {https://muse.jhu.edu/article/886933},
	doi = {10.1353/jod.2023.0016},
	language = {en},
	number = {2},
	urldate = {2025-03-01},
	journal = {Journal of Democracy},
	author = {Ganji, Sarath K.},
	month = apr,
	year = {2023},
	pages = {62--76},
}

@article{walsh_platform_2024,
	title = {Platform speech: {Journalists} and political campaigners reflect on {Facebook} and disintermediation in three {UK} general elections},
	issn = {1464-8849, 1741-3001},
	url = {https://journals.sagepub.com/doi/10.1177/14648849241273619},
	doi = {10.1177/14648849241273619},
	language = {en},
	urldate = {2025-03-08},
	journal = {Journalism},
	author = {Walsh, Matt and Singer, Jane B},
	month = sep,
	year = {2024},
	pages = {14648849241273619},
}
\end{document}